\documentclass[10pt]{iopart}
\pdfoutput=1

\expandafter\let\csname equation*\endcsname\relax

\expandafter\let\csname endequation*\endcsname\relax

\usepackage{amsmath}

\usepackage[utf8]{inputenc}
\usepackage{multirow}

\usepackage[dvips]{graphics,graphicx}
\usepackage{url}
\usepackage{subfigure}
\usepackage{etoolbox}
\usepackage{booktabs}
\usepackage[table,xcdraw]{xcolor}
\usepackage{ulem}
\usepackage{overpic}
\usepackage{hyperref}
\usepackage{amssymb}
\usepackage{color}
\usepackage{tablefootnote}

\renewcommand*{\emph}{\textit}

\newtoggle{draft}

\toggletrue{draft}
\newcommand{\g}{\cellcolor[HTML]{EFEFEF}}

\begin{document}
\title[Assessing energy dependence of the transport of relativistic electrons]{Assessing energy dependence of the transport of relativistic electrons in perturbed magnetic fields with orbit-following simulations}

\author{Konsta S\"arkim\"aki\textsuperscript{1}, Ola Embreus\textsuperscript{1}, Eric Nardon\textsuperscript{2}, T\"unde F\"ul\"op\textsuperscript{1}, and JET Contributors\textsuperscript{*}}

\address{\textsuperscript{1}
Department of Physics, Chalmers University of Technology, SE-41296 G\"oteborg, Sweden}
\address{\textsuperscript{2}
Association EURATOM-CEA, IRFM, CEA Cadarache, 13108 St-Paul-lez-Durance, France}
\address{\textsuperscript{*}
See the author list of E. Joffrin et al. 2019 Nucl. Fusion 59 112021}

\ead{konsta.sarkimaki@chalmers.se}
\vspace{10pt}
\begin{indented}
\item[]\today
\end{indented}

\begin{abstract}
%Experiments have shown that the theoretical estimate for the runaway electron transport can be several orders of magnitude higher than what is measured. 
%The discrepancy is attributed to transport decreasing for more energetic electrons as the finite orbit-width (FOW) effects become more relevant. 
%We show that that the theoretically predicted FOW-effects agree well with orbit-following simulations in perturbed tokamak magnetic fields. 
%Simulations are also extended to magnetic configurations characteristic of JET disruptions and to an ITER field perturbed with ELM control coils. 
%Our results show that FOW effects decrease transport only for the ITER case.  In contrast, for the studied JET cases we find that the presence of magnetic islands and nonuniform magnetic perturbations to have more dominant effects than the FOW effects. 
  %The diffusive-advective transport coefficients calculated in this work, based on simulations for electron energies 10 keV -- 100 MeV, can be used in reduced kinetic models to account for the transport due to the magnetic field perturbations.
  %Magnetic perturbations in disruptions can reduce runaway electron generation through increasing their radial transport.
Experimental observations, as well as theoretical predictions, indicate that the transport of energetic electrons decreases with energy. This reduction in transport is attributed to finite orbit width (FOW) effects. Using orbit-following simulations in perturbed tokamak magnetic fields that have an ideal homogeneous stochastic layer at the edge, we quantify the energy dependence of energetic electrons transport and confirm previous theoretical estimates. However, using magnetic configurations characteristic of JET disruptions, we find no reduction in RE transport at higher energies, which we attribute to the mode widths being comparable to the minor radius, making the FOW effects negligible. Instead, the presence of islands and nonuniform magnetic perturbations are found to be more important. The diffusive-advective transport coefficients calculated in this work, based on simulations for electron energies 10 keV -- 100 MeV, can be used in reduced kinetic models to account for the transport due to the magnetic field perturbations.
\end{abstract}

% Uncomment for PACS numbers
%\pacs{00.00, 20.00, 42.10}
%
% Uncomment for keywords
\vspace{2pc}

%
% Uncomment for Submitted to journal title message

% 
% Uncomment if a separate title page is required
%\maketitle
%
% For two-column output uncomment the next line and choose [10pt] rather than [12pt] in the \documentclass declaration
\ioptwocol
%

% Island trapping? https://www.osti.gov/servlets/purl/7034312
% On diffusion coefficient https://journals.aps.org/pre/abstract/10.1103/PhysRevE.73.026404

\section{Introduction}
\label{sec:Introduction}
Runaway electrons (REs) generated in disruptions are a major concern for future tokamaks with large plasma currents, such as ITER \cite{Lehnen_2015}. Predictions show that a large fraction of the initial plasma current can be converted into a runaway electron beam with energies of tens of MeV due to the avalanche process \cite{MHD_1999}. The subsequent uncontrolled loss of the RE current could damage the plasma facing components and has to be avoided. %The most discussed mitigation method is massive material injection \cite{HollmannDMS}. However, recent results suggest that injecting large amounts of material might even aggravate the runaway generation \cite{Hesslow_2019}.

The growth of the runaway electron population can be hindered through mechanisms that lead to radial losses. Substantial radial losses can reduce both the seed runaway population and the avalanche growth rate significantly~\cite{Helander_2000}.

Losses due to magnetic perturbations are expected to be present naturally in the early phase of the disruption, when the topology of the magnetic field confining the particles undergoes drastic changes, including the formation of stochastic regions leading to rapid radial transport of the REs.  Magnetic field perturbations can also be induced by external field coils. Such perturbations have been tested in several tokamaks, however, while runaways could be suppressed in medium-size tokamaks~\cite{Yoshino_2000,Lehnen_2008,Lin_2019}, the perturbations showed no significant effect on runaways on JET~\cite{Riccardo_2010}. The main focus of this paper is to quantify the radial losses of energetic electrons caused by magnetic perturbations in disruption scenarios.

Charged particle transport in a toroidal magnetic geometry with broken flux surfaces is often assumed to be diffusive with a diffusion coefficient that scales quadratically with the radial magnetic field perturbation amplitude $\delta B$~\cite{Rechester_1978,Spatschek_2008}:
\begin{equation}
\label{eq: rechester rosenbluth}
D \approx \frac{1}{\sqrt{2}}v_\parallel\lambda_\parallel\left( \frac{\delta B}{B}\right)^2.
\end{equation}
This type of diffusion arises when particles are moving along the stochastic field lines with velocity $v_\parallel$ and their orbits become decorrelated after travelling a distance $\lambda_\parallel$, the so-called parallel correlation length of the magnetic field.

%This collisionless transport mechanism is particularly relevant for disruption-born runaway electrons.
%Sufficient field stochasticity, either inherently present e.g. during the thermal quench or introduced via external magnetic coils, leads to increased transport and reduced avalanche growth rate, thus potentially mitigating the deleterious impact of the RE beam.

However, Eq.~\eqref{eq: rechester rosenbluth} has been found to overestimate the transport of REs (see the discussion in Ref.~\cite{Hauff_2009}). 
Two possible explanations have been put forward: the energy dependence of the transport is different from that implied by Eq.~\eqref{eq: rechester rosenbluth}, such that the transport is reduced for more energetic particles, or, the magnetic field is not completely ergodic and the remaining magnetic islands hinder the transport \cite{Hegna_1993}.  In this work we focus on the former issue---the energy dependence of the transport.

The non-trivial energy dependency of the diffusion coefficient is introduced via a scaling factor, $\Upsilon$,~\cite{Hauff_2009,Myra_1992} as
\begin{equation}
\label{eq: energy scaling}
D \approx \frac{1}{\sqrt{2}}v_\parallel\lambda_\parallel\left( \frac{\delta B}{B}\right)^2 \Upsilon.
\end{equation}
Physically, the scaling factor accounts for the finite orbit width (FOW) effects that arise when energetic particles drift from their initial flux surface, which interfere with the decorrelation process and lead to reduced transport, $\Upsilon\leq1$.
To clarify, FOW effects refer solely to the processes that are present even in a stochastic field with uniform structure; in a nonuniform field there could be additional mechanisms that are energy dependent e.g.~electrons with high energy being able to make excursions to a non-ergodic region. The existing theoretical work does not factor these other effects in $\Upsilon$.

The exact form and value of $\Upsilon$ depends on the magnetic perturbations via the parallel and the perpendicular magnetic field correlation lengths, $\lambda_\parallel$ and $\lambda_\perp$, and on the electron energy through the gyroradius and the orbit width:
\begin{align}
\rho_\mathrm{g} &= \frac{\gamma m v_\perp}{eB},\\
d_\mathrm{orb} &= q\frac{\gamma mv}{eB}.
\end{align}
Here $q$ is the safety factor and $\gamma$ is the Lorentz factor.
For electrons, FOW effects are expected to become important only at relativistic energies, $\gamma \gg 1$.
The explicit form of $\Upsilon$ is discussed in section~\ref{sec:theory} of this paper.

Note that the scenarios that were studied in Refs.~\cite{Hauff_2009,Esposito_1996,Entrop_1998} were turbulent flat-top scenarios, i.e.~the level of magnetic perturbations was considerably lower than the one typical of the thermal quench of a disruptive plasma. Although the theoretical estimates relevant to flat-top scenarios are not expected to hold in disruptions, they can be used to benchmark the numerical models.

The energy dependence of the RE transport can also be studied numerically using orbit-following simulations.
The benefit of such simulations is that they can accurately model motion of particles and their transport due to magnetic perturbations using realistic magnetic fields, if those can be provided.
In this work, we use orbit-following simulations to:
\begin{itemize}
\item Assess whether significant reduction in RE transport occurs (i.e. $\Upsilon<1$) at high energy in disruption magnetic fields, or when external coils are used for RE mitigation, for plausible RE energies (section~\ref{sec:simulations}).
\item Verify the theoretical estimate for the energy-dependence of the transport given in Ref.~\cite{Hauff_2009} (section~\ref{sec:comparison}).
\end{itemize}
We quantify the cross-field transport of REs by computing the corresponding (energy-dependent) advection and diffusion coefficients.
The advantage of this method is that the transport coefficients enable one to introduce 3D magnetic field effects to reduced kinetic models, which makes it possible to assess the reduction in avalanche growth rate due to the field stochasticity.
However, this task is beyond the scope of the present work.

\begin{figure}[t]
\centering
\begin{overpic}[width=0.45\textwidth]{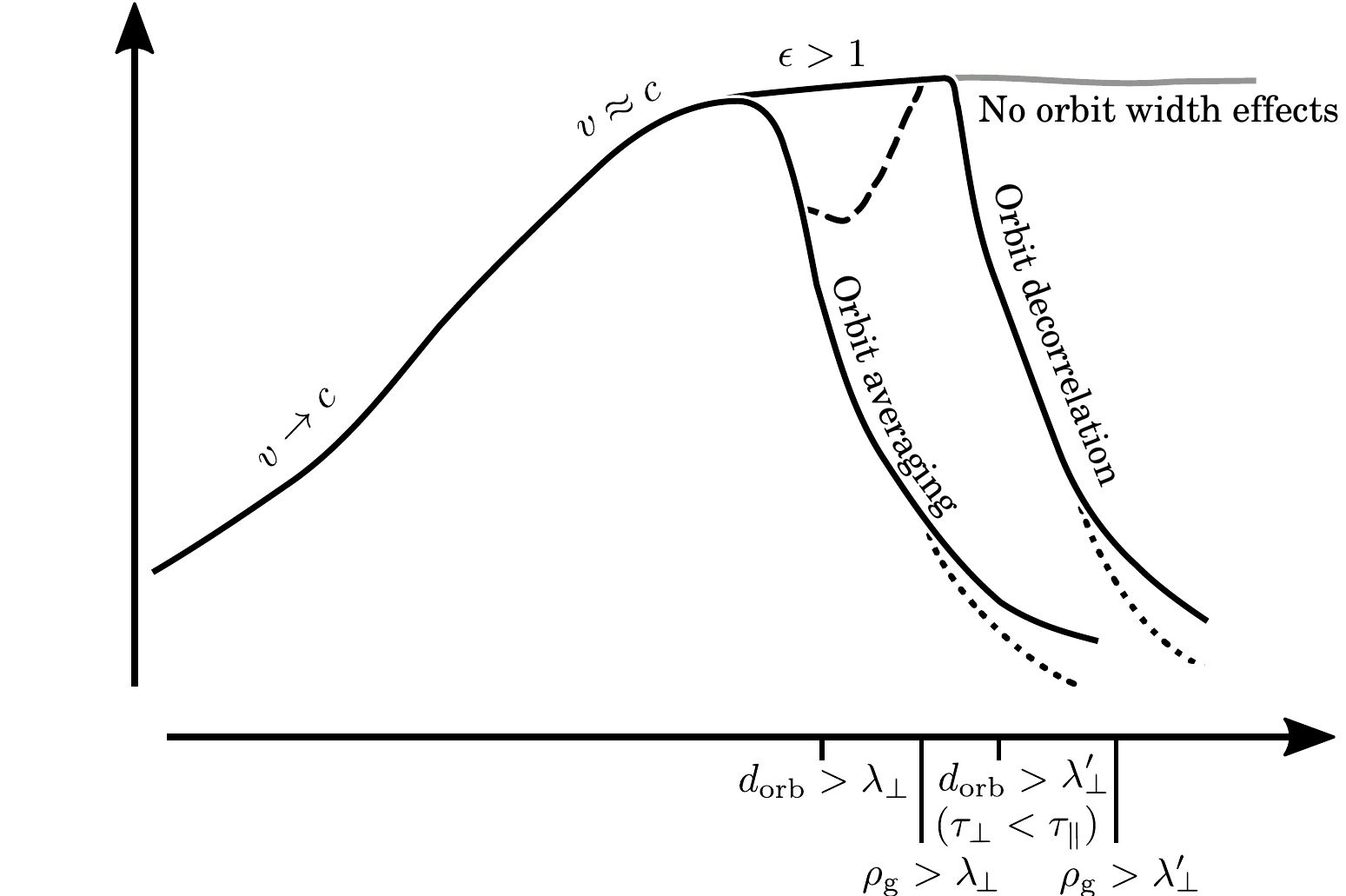}
\put(20,6){$\log$ Energy}
\put(3,30){\rotatebox{90}{Transport}}
\end{overpic}
\caption{
Illustration of the expected energy dependence of the transport.
Here $\lambda_\perp' > \lambda$, and the conditions $d_\mathrm{orb}>\lambda_\perp'$ and $\tau_\perp<\tau_\parallel$ are met simultaneously.
The dotted lines include finite gyro-radius effects.
}
\label{fig: sketch energydep}
\end{figure}

\section{Transport in the presence of finite orbit-width effects}
\label{sec:theory}

Figure~\ref{fig: sketch energydep} illustrates the expected energy dependence of the radial transport. 
After increasing initially due to increasing velocity, the transport ceases to grow when $v\approx c$, where it would remain flat if FOW effects were not taken into account.
At higher energy, the exact behaviour depends on two mechanisms, \emph{orbit-averaging} and \emph{perpendicular decorrelation}, that are responsible for FOW effects.
These mechanisms are presented in detail in Ref.~\cite{Hauff_2009} and here we only review the main points.

Particles with finite orbits do not trace the same field line exactly due to the radial drift.
This oscillation in radius is accompanied by toroidal precession, which is the transit-time averaged drift toroidally.
While the particle returns to same radial position after each transit, it moves toroidally from the original field line by a distance equal to $x=v_\mathrm{prec}\tau_\mathrm{orb}$, where
\begin{equation}
v_\mathrm{prec} = \frac{\gamma mv^2\hat{s}}{eBR_0} ,
\end{equation}
is the toroidal precession velocity, $R_0$ the major radius, $\hat{s}$ the magnetic shear, and
\begin{equation}
\tau_\mathrm{orb} = \frac{2\pi qR_0}{v},
\end{equation}
is the transit time.
The particle returns to the zone where it still remains correlated with the original perturbation if
\begin{equation}
\label{eq: orbit decorrelation param}
\epsilon \equiv \frac{v_\mathrm{prec}\tau_\mathrm{orb}}{\lambda_\perp} < 1.
\end{equation}
The parameter $\epsilon$ is referred as the \emph{orbit-averaging validity parameter}.

When the particle returns to (or does not leave at all) the zone of correlation, $\epsilon<1$, the transport decreases because the particle experiences effectively smaller perturbation as the fluctuating magnetic perturbation is averaged along the particle trajectory.
This effect becomes more prominent as the orbit width becomes comparable to the perpendicular correlation length, eventually leading to a $\Upsilon \propto \gamma^{-1}$ scaling in transport when $d_\mathrm{orb} \gtrsim \lambda_\perp$.
The explicit formulae for $\Upsilon$ are given in Ref.~\cite{Hauff_2009}
\begin{align}
\label{eq: orb avg small d}
\Upsilon &\approx \left[1-2\left(\frac{d_\mathrm{orb}}{\lambda_\perp}\right)^2
+\frac{5}{2}\left(\frac{d_\mathrm{orb}}{\lambda_\perp}\right)^4\right]^2&\forall d_\mathrm{orb} \lesssim \lambda_\perp,\\
\label{eq: orb avg large d}
\Upsilon &= \frac{\lambda_\perp}{ d_\mathrm{orb}} &\forall d_\mathrm{orb} \gtrsim \lambda_\perp.
\end{align}
If the electron energy is high enough, the orbit-averaging also happens on the gyromotion scale if $\rho_g  \gtrsim 0.3\lambda_\perp$. 
This additional scaling is obtained from Eqs.~\eqref{eq: orb avg small d}~--~\eqref{eq: orb avg large d} when $d_\mathrm{orb}$ is replaced with $\rho_g$.
Since the perturbation is averaged both along the poloidal and the gyro orbit, eventually the transport decreases as $\Upsilon \propto \gamma^{-2}$.

If the particle does not return to the zone of correlation, $\epsilon>1$, the orbit-averaging mechanism is not valid.
However, the particle becomes decorrelated during the radial excursion if the orbit width is wide enough, $d_\mathrm{orb} > \lambda_\perp$, and the characteristic time-scale for this perpendicular decorrelation
\begin{equation}
\label{eq: perp decorr time}
\tau_\perp \equiv \frac{\tau_\mathrm{orb}\lambda_\perp}{2\pi d_\mathrm{orb}},
\end{equation}
is smaller than that of the parallel decorrelation
\begin{equation}
\label{eq: parallel decorr time}
\tau_\parallel \equiv \frac{\lambda_\parallel}{v_\parallel},
\end{equation}
i.e. $\tau_\perp < \tau_\parallel$.
The energy scaling for this process is given in Ref.~\cite{Hauff_2009}
\begin{equation}
\label{eq: orb decor large d}
\Upsilon = \frac{\lambda_\perp}{\lambda_\parallel d_\mathrm{orb}} \qquad\forall\; d_\mathrm{orb} > \lambda_\perp.
\end{equation}
The case $d_\mathrm{orb} < \lambda_\perp$ is not covered in the literature, but we expect that the transport to remain constant in that case or it might increase to its original level if it was reduced before orbit-averaging becomes invalid (the dashed line in Fig.~\ref{fig: sketch energydep}).
The orbit-averaging is still valid for the gyromotion when $\rho_g \gtrsim 0.3\lambda_\perp$ and, thus, one again observes $\Upsilon \propto \gamma^{-2}$ at sufficiently high energies.

These scaling laws are based on the small \emph{Kubo number} approximation, $\mathcal{K}\equiv (\lambda_\parallel / \lambda_\perp)(\delta B / B) < 1$.
However, this is also the condition for the validity of the quasilinear approximation, Eq.~\eqref{eq: rechester rosenbluth}, for the particle transport to be valid, and so we restrict this study to the $\mathcal{K}<1$ regime.

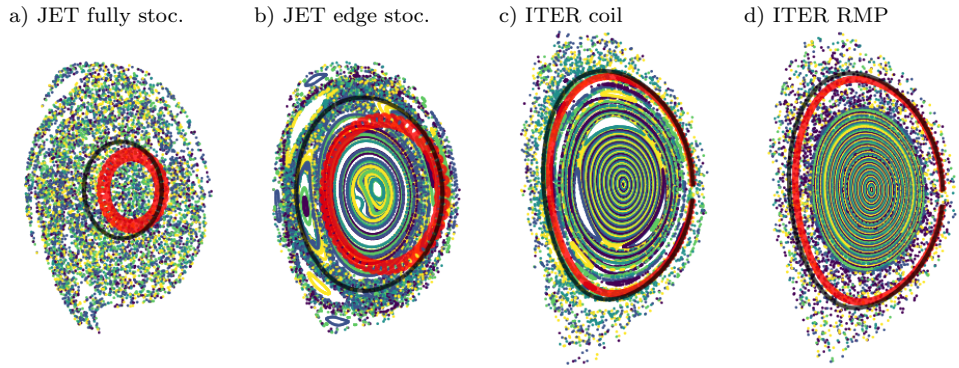
\begin{figure*}[t]
\centering
\begin{overpic}[width=0.95\textwidth]{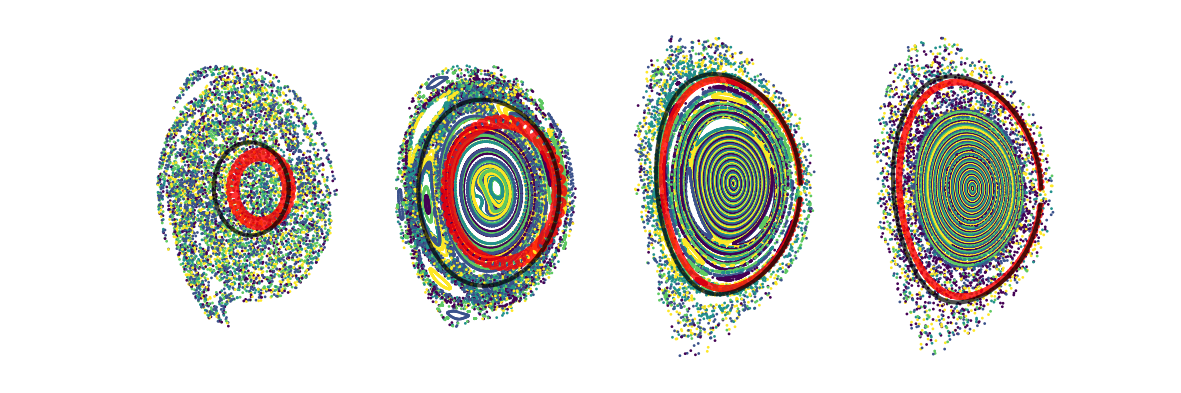}
\put(12,31){\footnotesize{a) JET fully stoc.}}
\put(32,31){\footnotesize{b) JET edge stoc.}}
\put(52,31){\footnotesize{c) ITER coil}}
\put(72,31){\footnotesize{d) ITER RMP}}
\end{overpic}
\caption{
Magnetic field Poincar\'e-plot at $\phi=0^\circ$ $Rz$-plane for the investigated cases.
\textbf{a)} Fully stochastic JET magnetic field during the thermal quench.
\textbf{b)} Partially stochastic JET magnetic field during the pre-thermal quench phase.
\textbf{c)} ITER current flat-top equilibrium perturbed with ELM control coils.
\textbf{d)} ITER current flat-top equilibrium perturbed with artificial resonant magnetic perturbations.
The red and black orbits illustrate the trajectories of 100 MeV and 10 keV electrons which are the extreme values used in the simulations. 
On each plot, the trajectories begin at the outer mid-plane at the same radial position where the markers where initialized in the actual simulations.
}
\label{fig: poincares}
\end{figure*}

\section{Energy scan in perturbed JET and ITER fields}
\label{sec:simulations}

\subsection{Magnetic backgrounds}

Even though the transport is expected to decrease at sufficiently high energy, a more interesting point is to know whether this happens within plausible range of RE energies.
The answer to this question depends on the perpendicular correlation length $\lambda_\perp$, which dictates the scaling laws and whether the orbit-averaging is valid. 
The value of $\lambda_\perp$ depends on the magnetic field structure and the nature of the perturbation, making it impossible to give a universal answer.
Instead, here we assess the energy dependence of the transport by investigating three cases which are relevant for RE dynamics and mitigation in disruptions, each representing a different type of magnetic configuration.

The first case represents a magnetic field that has become fully stochastic, as shown in Fig.~\ref{fig: poincares}a, during the thermal quench phase in a plasma disruption.
The second case is during the pre-thermal quench phase, when the core has not yet become stochastic and some magnetic islands remain at the stochastic edge region (Fig.~\ref{fig: poincares}b).
These two cases are important because it is during the thermal quench, when the initial seed population for the RE avalanche is born.
During the current quench phase following the complete stochastization, the flux surfaces begin to heal, which can be expected to lead to better RE confinement.
%Unfortunately, no data was available to simulate transport in this phase. \cmt{this last sentence is irrelevant - and not even true, we could have asked for that data; I do not delete it but it should be deleted.}

The magnetic field data for the previous two cases are obtained from a JOREK 3D non-linear MHD simulation \cite{jorek1,jorek2}. The simulated case is a massive injection of Argon gas in JET pulse number 85943, which led to the formation of a RE current of 1 MA. Details of the model will be described in a forthcoming publication \cite{Hu2020}. This simulation has been chosen in particular because a plasma current ($I_p$) spike of comparable magnitude to the experimental one is obtained, which we interpret as a likely sign that the level of magnetic stochasticity is well reproduced. Indeed, theoretical work \cite{Boozer_PPCF_2018} as well as JOREK simulations indicate a correlation between the $I_p$ spike height and the level of magnetic stochasticity.

The third case (Fig.~\ref{fig: poincares}c) corresponds to a proposed scenario for RE mitigation, where the field is made stochastic using external coils.
In ITER, the ELM control coils could be used for this purpose, although they are believed to be insufficient for RE mitigation on their own \cite{Papp_2011PPCF,Papp_2012PPCF}.
Nevertheless, we use the current flat-top magnetic field (no post-disruption data is available) of the ITER baseline scenario perturbed with ELM control coils for the transport study.
The field is similar and constructed with same means as to what was used in an earlier study~\cite{Sarkimaki_2016} that did not investigate FOW effects.
The coils are set to an $N=3$ mode, where $N$ is the toroidal periodicity, with coil current $I=45$ kA which is half the maximum value (this is the setup foreseen to be used for ELM mitigation~\cite{evans20133d}).
The field is constructed by adding the perturbation due to the coils, calculated with a Biot-Savart solver~\cite{akaslompolo2015biot}, on the equilibrium field.
This results in a field where a stochastic layer is generated at the edge but leaves the core intact.

In addition to these three realistic cases, we have constructed an artificial field by imposing arbitrary resonant magnetic perturbations to the ITER equilibrium.
These perturbations create a stochastic layer at the edge, with no islands, as shown in Fig.~\ref{fig: poincares}d, for which the quantities of interest, namely $\delta B / B$, $\lambda_\parallel$, and $\lambda_\perp$, are adjustable.
The perturbations are stationary and have the form
\begin{equation}
\label{eq: nabla alpha B}
\delta\mathbf{B} = \delta_0\nabla\times\left(\alpha\mathbf{B}_{2D}\right)
\end{equation}
where $\mathbf{B}_{2D}$ is the (unperturbed) axisymmetric field, $\delta_0$ a perturbation scaling factor, and
\begin{equation}
\label{eq: helical perturbation}
\alpha(\rho,\;\theta,\;\zeta) = \sum_{n,m}\alpha_{nm}(\rho)\cos(n\zeta - m\theta - \phi_{0,nm}).
\end{equation}
Here $(\rho,\; \theta,\; \zeta)$ are the radial, poloidal, and toroidal straight-field-line coordinates, respectively.
Each mode is defined by its radial eigenfunction $\alpha_{nm}$, poloidal, $m$, and toroidal, $n$, mode numbers, and phase $\phi_{0,nm}$, which is chosen to be random.

\begin{figure}[ht]
\centering
\begin{overpic}[width=0.45\textwidth]{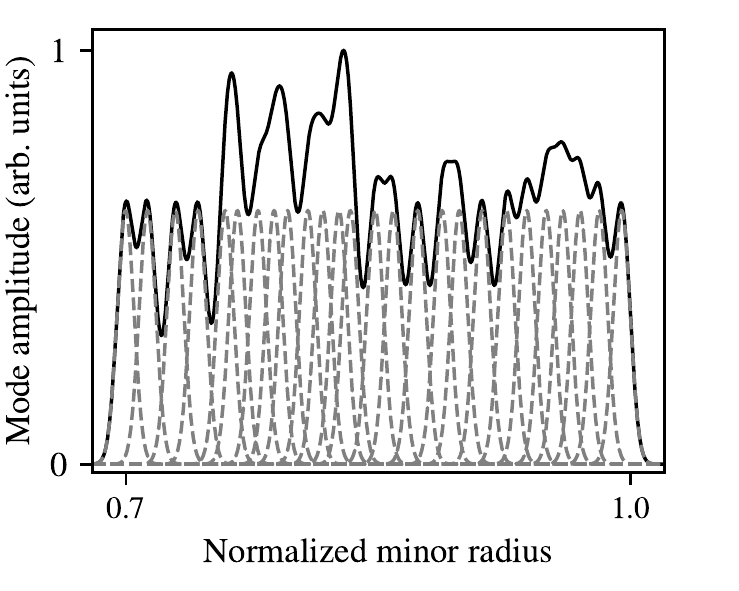}
\end{overpic}
\caption{
Radial distribution of the mode eigenfunctions in ITER RMP simulations.
Each mode has an eigenfunction which is a Gaussian whose peak is located on the mode's rational surface.
The modes (dashed lines) are distributed radially almost evenly.
The solid line shows the total perturbation.
}
\label{fig: modes}
\end{figure}

The eigenfunctions are Gaussian functions that peaks at the corresponding resonant surface, $\alpha_{nm}(\rho)=\exp(-(\rho-\rho_{nm})^2/2\sigma^2)$ where $q(\rho_{nm}) = m/n$, and the width $\sigma$ is left as a free parameter.
A total of 25 modes were used to create the stochastic layer at the edge, and the mode numbers $n$ and $m$ were chosen so that the resonant surfaces were distributed at approximately equal intervals in $\rho$ (see Fig.~\ref{fig: modes}).
The scaling factor $\delta_0$ is chosen so that $\delta B / B \approx 1\times10^{-3}$.
Note that Ref.~\cite{Myra_1992}, where the orbit-averaging process was originally presented, used similar perturbations to study the transport in perturbed magnetic fields.

\subsection{Transport evaluation}

\begin{figure*}[t]
\centering
\begin{overpic}[width=0.95\textwidth]{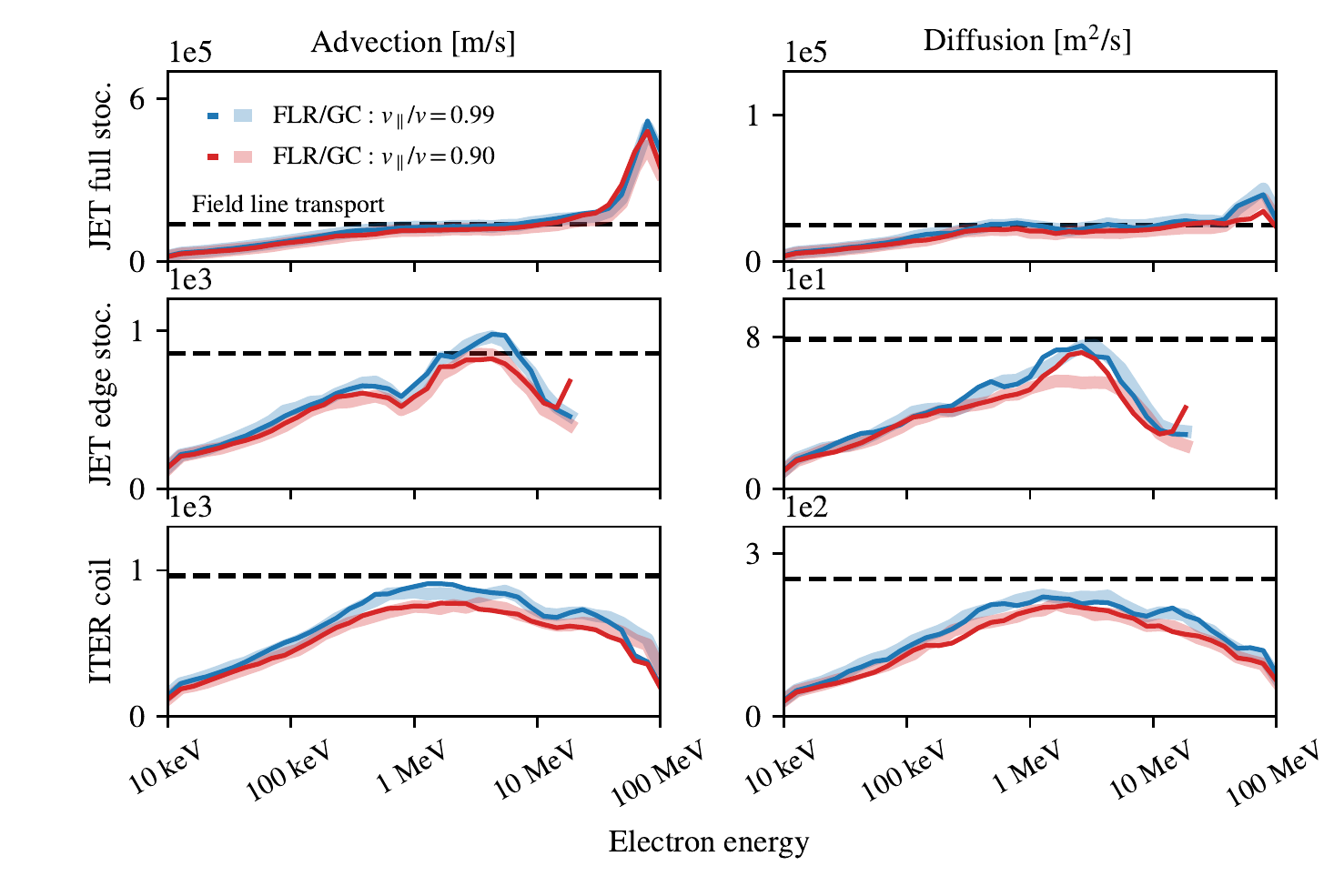}
\end{overpic}
\caption{
Transport coefficients evaluated with orbit-following simulations.
Each row corresponds to a different case.
The thin bright lines correspond to gyro-orbit simulations, i.e., the ones including finite Larmor radius effects.
The thick faded lines correspond to the guiding-center results.
Blue color corresponds to particles with pitch $v_\parallel/v=0.99$ and red color to $v_\parallel/v=0.90$.
Each line is drawn from a simulation with 32 distinct energies, and the Monte Carlo noise was reduced by taking a moving average before plotting.
The dashed black line is the magnitude of the field-line transport coefficient, i.e., transport of massless particles travelling at the speed of light along the magnetic field.
}
\label{fig: transport coeff}
\end{figure*}

It was found in Refs.~\cite{Sarkimaki_2016,Papp_2015}, that modelling the transport as a purely diffusive process can be inaccurate.
Better correspondence between the actual and the modelled transport is obtained by also including an advection component to the transport model~\cite{Sarkimaki_2016} (that is in addition to the effective advection caused by the gradient in the diffusion coefficient).
However, we emphasize that the choice of modelling the transport as an advection-diffusion process was motivated by its numerical practicality for implementation in reduced models and the convenience in evaluating the transport coefficients.
%For a more fundamental approach, Refs.~\cmt{(add refs)} give a good overview on the nature of the transport. \cmt{I deleted this because it is not needed for the storyline, it just disturbes the flow and diverts attention}

The transport coefficients are evaluated with the Monte Carlo method from several markers, representing electrons, which are traced in a perturbed magnetic field using the orbit-following code ASCOT5~\cite{ascot5}.
The transport coefficients are evaluated by simulating a population of markers that were initially located on the same radial position in the stochastic region.
The initial position in each case is displayed in Fig.~\ref{fig: poincares}.
The markers were simulated until the time distribution of the cumulative losses saturated, and the distribution was subsequently used to extract the transport coefficients.
In effect, the coefficients obtained via this process do not describe transport locally, but they can be considered to describe the average transport within the entire stochastic region.
The details of obtaining transport coefficients from the orbit-following simulation results are explained in \ref{app:a}.

In these simulations, all transport will be due to magnetic field perturbations, as the electrons are assumed to be collisionless and the radiation reaction force is ignored. In addition, also the acceleration due to electric field is omitted so the electron energy remains constant, allowing us to scan the transport for different values of electron energy.

Since we expect to see a reduction in transport not only due to finite poloidal orbit width effects but also due to finite Larmor radius (FLR), each setting is simulated twice:
once solving for the full gyro-motion, and another time solving for the guiding center motion. 
The FLR effects can be isolated by comparing the results.

The simulations are done for different energies ranging from 10 keV to 100 MeV.
The purpose is to observe whether the expected energy dependence (recall Fig.~\ref{fig: sketch energydep}) can be recovered in magnetic fields relevant for RE mitigation.
Furthermore, we carry out additional simulations where the field lines themselves are traced instead of electrons, in order to obtain the zero-orbit-width results for comparison.

\subsection{Results}

The transport coefficients evaluated using the orbit-following simulations are shown in Fig.~\ref{fig: transport coeff} for the three cases with realistic magnetic field.
The transport in the ITER field with artificial perturbations will be discussed separately in the next section.
The results show good agreement between the guiding-center and gyro-orbit simulations, indicating that the FLR effects play only a minor role in the studied cases.
The transport does depend on pitch as the results with $\xi\equiv v_\parallel/v=0.90$ are generally lower than the ones with $\xi=0.99$. However, the difference is in line with the trivial pitch dependence, $v_\parallel \propto \xi$ which is present even in the Rechester-Rosenbluth diffusion coefficient given in Eq.~\eqref{eq: rechester rosenbluth}.
The energy dependence of the transport shows the expected behavior for low energies ($E<1$ MeV) where the transport is proportional to $v_\parallel$.
When $v \approx c$, the transport either peaks or flattens at the level of the field-line transport.
For higher energies, the behavior differs in each case.

\emph{The first row:} The fully stochastic magnetic field case has both advection and diffusion rising above the field-line transport level at $E>10$ MeV.
This result is inconsistent with with the hypothesis that $\Upsilon \leq 1$ in Eq.~\eqref{eq: energy scaling}.
A possible explanation for the increased transport at higher energies is that there exists a transport barrier between the edge and the initial location of the particles.
More energetic electrons would be more able to cross this barrier if the barrier width is less than the electron orbit width.
We will show later on that such a barrier exists, and the observed $\Upsilon > 1$ can be attributed to it.

%which is inconsistent with the hypothesis that $\Upsilon \leq 1$ in Eq.~\eqref{eq: energy scaling}, is because the advection-diffusion approach breaks down; the loss-time distributions show that the transport becomes more, but not entirely, laminar for higher energies.
%A possible explanation is that because $\delta B/B$ varies strongly along the poloidal trajectory of a particle, with the perturbation being strongest near the core, the resulting gradient increases advection to the point where the transport turns to laminar.
%Another possibility is that there exists a radial region with considerably lower transport which particles have to cross before they are lost.
%If the width of this layer is comparable to the electron orbit width or gyroradius, then the more energetic electrons are more readily to cross it.

\emph{The second row:} The edge stochasticity case has a drop in both advection and diffusion above a few MeV.
There is a simultaneous decrease in the number of markers that are lost within the simulation time.
For energies above 20 MeV, the few markers which are lost, do so in groups: the transport has become completely laminar and the coefficients cannot be sensibly evaluated, hence the abruptly ending lines. These effects are caused by the magnetic field topology: the electrons with high energy cross the region where the flux surfaces are intact, and no stochastic field line transport can occur.
For example, 100 MeV electrons spend a major part of their orbit within the region with intact flux surfaces, as was illustrated in Fig.~\ref{fig: poincares}b.
The more time is spent within the region with intact flux surfaces, the larger the reduction in transport until it vanishes completely.
In other words, the observed reduction in transport is not due to FOW effects, such as orbit-averaging.
We can confirm this interpretation by moving the radial position outwards, where the markers are initialized,
This increases the energy threshold above which the transport begins to decrease.

\emph{The third row:} The ITER coil case has markers initialized at the very edge where there are no islands and even the $E=100$ MeV electrons cannot reach the intact flux surface region initially.
Hence, there is no similar cutoff in transport at higher energies as observed in the previous case.
However, the transport decreases with increasing energy, which could be attributed to FOW effects.
On the other hand, the slope indicates a reduction weaker than the predicted $\Upsilon \propto \gamma^{-1}$.

To summarize, none of the considered JET cases exhibit a reduction of transport with energy compatible with orbit averaging or orbit decorrelation effects. % the expected reduction in energy. 
In the fully stochastic case the transport \emph{increases} as a function of energy for reasons to be clarified.
In the case with stochastic edge, the transport decreases but this is due to particles spending more time in their excursion in the healed flux surface region.
Only the ITER coil case shows a reduction in transport with increased energy that could be attributable to FOW effects. 
However, the reduction of the transport at high energy is weaker than expected.

\begin{table*}[ht]
\caption{Magnetic field parameters and threshold energies.}
\label{tbl: field params}
\resizebox{2\columnwidth}{!}{
\begin{tabular}{@{}clcccccccccc@{}} \toprule
&  & \multicolumn{5}{c}{\textbf{Magnetic field parameters}}                                                                                                                                                                &  & \multicolumn{3}{c}{\textbf{Threshold energies [MeV]}}                                                                                                                                                                 \\
& & 
\multicolumn{1}{c}{$\lambda_\parallel$ [m]} & 
\multicolumn{1}{c}{$\lambda_\perp$ [m]} & 
\multicolumn{1}{c}{$\delta B/B$} & 
\multicolumn{1}{c}{$\mathcal{K}$} & 
\multicolumn{1}{c}{$D/D_\mathrm{num}$} & 
& 
\multicolumn{1}{c}{$\epsilon>1$} & 
\multicolumn{1}{c}{$d_\mathrm{orb}>\lambda_\perp$} & 
\multicolumn{1}{c}{$\rho_g>0.3\lambda_\perp$} &
\multicolumn{1}{c}{$\tau_\perp<\tau_\parallel$} \\ 
\cmidrule(lr){3-7} \cmidrule(l){9-12} 
\parbox[t]{2mm}{\multirow{6}{*}{\rotatebox[origin=c]{90}{$n=10$}}} 
& $\sigma=0.01$ m & 2 & 0.006 & $9\times10^{-4}$ & 0.3 & 0.6 & & 0.6 & 5 & 10 & 10\\
& & & & & & & & & & &\\
& $\sigma=0.03$ m\g & 7\g & 0.04\g & $7\times10^{-4}$\g & 0.1\g & 0.7\g & \g & 6\g & 30\g & 100\g & 30\g\\
& \g & \g & \g & \g & \g & \g & \g & \g & \g & \g & \g\\
& $\sigma=0.2$ m & 2 & 0.3 & $1\times10^{-3}$& 0.04 & 0.9 & & 20 & 90 & 200 & 100\\
& & & & & & & & & & &\\

\cmidrule(l){2-12} 

\multirow{6}{*}{} 
& Full stoc. & 8 & 0.3 & $7\times10^{-3}$ & 0.2 & 5 & & 50 & 100 & 200 & 50\\
& (JET)	 & & & & & & & & & &\\
& Edge stoc.\g & 5\g & 0.3\g & $2\times10^{-3}$\g & 0.03\g & 100\g & 
 \g & 50\g & 100\g & 200\g & 70\g\\
& (JET)     \g &  \g &    \g &                 \g &     \g & \g & 
 \g & \g & \g & \g & \g\\
& Coil       & 2 & 0.7 & $9\times10^{-4}$ & 0.003 & 1 & & 100 & 600 & $2\times10^3$ & $2\times10^3$\\
& (ITER) & & & & & & & & & &\\

\bottomrule
\end{tabular}
}
\end{table*}

\section{Comparison to analytical results}
\label{sec:comparison}

In order to better understand the results in the realistic cases, we now study the transport in the presence of artificial magnetic perturbations superimposed on the ITER equilibrium, allowing us to adjust the perturbation parameters at will.

Based on the discussion in section~\ref{sec:theory}, FOW effects become apparent when the perpendicular correlation length, $\lambda_\perp$, is comparable to the electron orbit width.
The orbit width for a 100~MeV electron in ITER is approximately 0.1~m.
To observe FOW effects over the energy range of 10~keV~--~100~MeV, we therefore seek to have $\lambda_\perp \approx$ 0.001~--~0.1~m ($d_\mathrm{orb} \propto \gamma$ when $\gamma \gg 1$).
We can make an intuitive guess that the perpendicular correlation length is comparable to the mode width, $\lambda_\perp \sim \sigma$, when the perturbations have Gaussian radial profiles.
For the parallel correlation length, we may use the estimate $\lambda_\parallel \approx 2\pi q R / n$.

We choose the toroidal and poloidal mode numbers as $n\lesssim$ and $m\lesssim 20$, so that the mode resonant surfaces are evenly spaced radially~\footnote{
The dominant modes in the realistic cases were $n\leq3$, but with $n=3$ there were not enough rational surfaces to enable several adjacent modes.
}.
In total there are 25 adjacent modes, and the smallest possible width that still allows the modes to overlap and keeps the edge region stochastic is $\sigma=0.01$~m.
We repeat the simulations for three cases where the mode width is varied from this smallest possible value to the width of the entire stochastic layer.

With these choices, we estimate that $\lambda_\parallel\approx 8$~m and $\lambda_\perp \approx 0.01$~--~0.2 m.
\ref{app:b} presents how the correlation lengths are computed numerically. In the following discussion, we rely on the numerically evaluated values.

\begin{figure*}[t]
\centering
\begin{overpic}[width=0.95\textwidth]{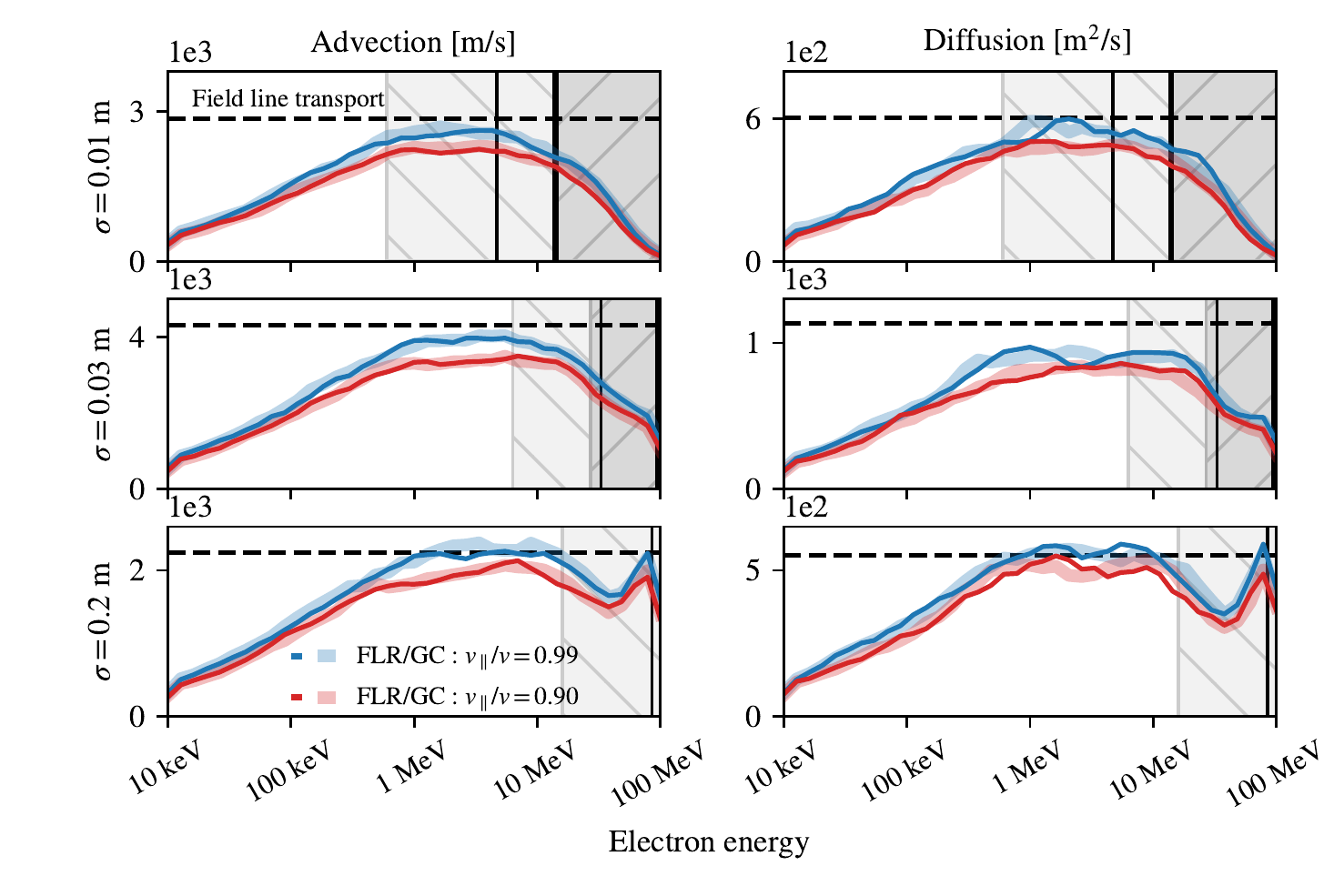}
\end{overpic}
\caption{
Transport coefficients evaluated with the orbit-following simulations for the ITER RMP case with $n=10$ and for different values of mode width $\sigma$.
The various grey regions in the background indicate when the orbit averaging has become invalid, $\epsilon>1$, (light grey with "\textbackslash{}"-stripes), and when $\tau_\perp<\tau_\parallel$ (dark grey with "/"-stripes).
The plain white background is where the orbit-averaging is valid.
The black vertical lines indicate when the threshold energies are reached:
thin lines indicate the $d_\mathrm{orb} > \lambda_\perp$ threshold whereas the thick lines are for the $\rho_\mathrm{g} > 0.3\lambda_\perp$ threshold (for the $v_\parallel/v=0.90$ case).
}
\label{fig: scan sigma}
\end{figure*}

The magnetic field parameters for each simulated case are listed in Table~\ref{tbl: field params}.
The first and second columns show the parallel and perpendicular correlation length, respectively, and the third column is the flux surface average of the perturbation magnitude on the initial radial position.
The fourth column is the Kubo number evaluated from the aforementioned parameters.
In all cases $\mathcal{K} \ll 1$ making the quasilinear approximation, Eq.~\eqref{eq: rechester rosenbluth}, valid.
One can confirm this by evaluating the ratio between the numerically evaluated field-line diffusion coefficient, $D_\mathrm{num}$, and the analytical estimate, Eq.~\eqref{eq: rechester rosenbluth}.
This ratio is tabulated in the fifth column, and it is close to unity in almost all cases, with the exception of the stochastic edge JET case, possibly because there are major magnetic islands present. 

Table~\ref{tbl: field params} also lists \emph{threshold energies}, corresponding to the following situations: the orbit-averaging becomes invalid, the orbit width exceeds the perpendicular correlation length, and when the orbit-averaging occurs on the gyromotion scale.
The final column lists the threshold energy for $\tau_\perp < \tau_\parallel$, the third condition for the perpendicular orbit decorrelation mechanism (the two others were $\epsilon>1$ and $d_\mathrm{orb} > \lambda_\perp$).

We first compare these threshold energies to the calculated transport coefficients in the artificial field (shown in Fig.~\ref{fig: scan sigma}) before returning to the discussion of the realistic cases studied earlier.
Each case in Fig.~\ref{fig: scan sigma} exhibits an increasing transport as $v\rightarrow c$, until it saturates and begins to decrease.
In each case $\Upsilon\leq 1$, as the theory predicts, but the exact behaviour of the transport varies between the cases.

\emph{The first row:} Orbit averaging becomes invalid right when $v\approx c$ is reached and no transport reduction is observed until $d_\mathrm{orb}>\lambda_\perp$.
Afterwards there is a slight reduction (when the transport should be constant), which could be caused by a transition in the mechanism driving the transport, from parallel decorrelation to perpendicular decorrelation as $\tau_\parallel\rightarrow\tau_\perp$.
The reduction in transport steepens as the perpendicular decorrelation and the gyro-orbit effects become relevant---by a coincidence, almost at the same energy.
Here we observe the $\gamma^{-1}$ relation between the transport and the electron energy.
However, even though the FLR effects should be present, only a small difference is observed when those are accounted for.

\emph{The second row:} The orbit-averaging is valid for higher energies than in the previous case.
Equation~\eqref{eq: orb avg small d} predicts a maximum of 10\% reduction until the orbit-averaging becomes invalid.
However, no reduction is observed until the point where the perpendicular orbit-decorrelation becomes valid.
At the very end of the energy range the FLR effects become valid, and there it appears that the transport begins to decrease more steeply.

\emph{The third row:} In this case, the orbit averaging is valid for even higher energies, but the notable feature is the valley near $E=50$ MeV.
If the orbit-averaging would be valid at that point, the reduction in transport would be 15\%, which is close to the observed value.
We suggest that the increase beyond this point is due to the orbit-averaging transitionally becoming invalid and the transport returning to $\Upsilon=1$.
In other words, what is seen here corresponds to the dashed line in Fig.~\ref{fig: sketch energydep}.
At the end of the energy range, the transport peaks and begins to decrease again.
This is at the threshold when the perpendicular decorrelation becomes valid.

In summary, we observe a good, though not an exact, agreement between the theoretical estimates and the simulations.
The differences are mostly related to the orbit-averaging process.

\section{Realistic cases revisited}
\label{sec:discussion}

Having verified the theory through the simulations, and vice versa, we can now compare the transport in the realistic cases  to the theoretical predictions.

In the JET cases, recall Fig.~\ref{fig: transport coeff}, the orbit averaging is valid until $E=50$ MeV.
At this energy, where the electron orbit width is approximately 10~cm, Eq.~\eqref{eq: orb avg small d} predicts a reduction of 35\% in transport.
However, this reduction is not visible in the results due to the transport barrier effect being dominant in the fully stochastic case and the more energetic electrons becoming confined in the case with the stochastic edge.
The perpendicular decorrelation mechanism becomes active at 100~MeV, when $d_\mathrm{orb} > \lambda_\perp$, and one can observe a drop in transport in the first row in Fig.~\ref{fig: transport coeff} that could be attributed to this.

%In the JET cases, recall Fig.~\ref{fig: transport coeff}, the orbit averaging is valid until $E=50$ MeV.
%However, the other effects, i.e.~those caused by the transport barrier and the intact flux surfaces, have already become dominant at this point.
%The orbit width of a 10~MeV electron, for which we were still able to evaluate the coefficients before markers became confined in the case with non-ergodic regions (the second row in Fig.~\ref{fig: transport coeff}), is approximately 2 cm.
%At this energy, Eq.~\eqref{eq: orb avg small d} predicts a reduction of 2\% in the transport.
%At the first row in Fig.~\ref{fig: transport coeff}, the decreasing transport at the end of the energy range could be attributable to the perpendicular decorrelation mechanism, which activates at 100~MeV when $d_\mathrm{orb} > \lambda_\perp$.

\begin{figure}[t]
\centering
\begin{overpic}[width=0.45\textwidth]{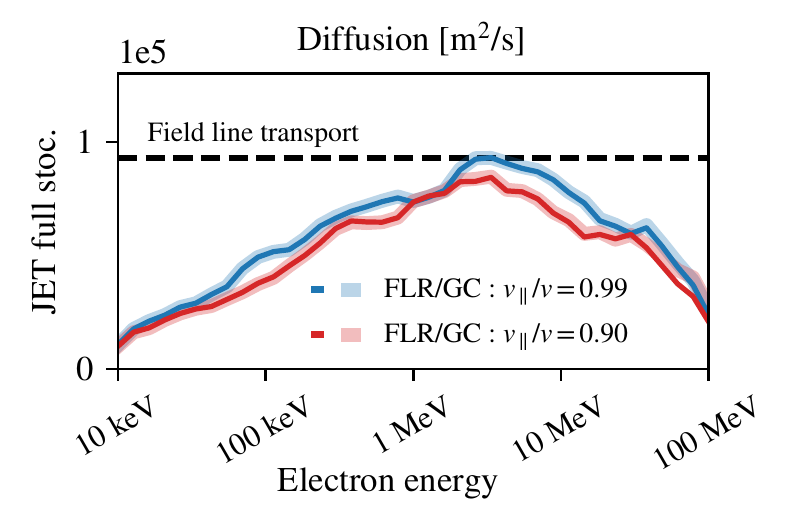}
\end{overpic}
\caption{
Diffusion coefficient in the fully stochastic JET case when the loss surface is moved inwards over the transport barrier.
}
\label{fig: newdiffusion}
\end{figure}

We can confirm the presence of the transport barrier at the edge in the fully stochastic JET case by moving inwards the loss surface, i.e., the boundary beyond which markers are considered lost.
Figure~\ref{fig: newdiffusion} shows the energy dependence of the transport when the loss surface has been moved over the barrier.
Now the reduction in transport is similar to what the theory predicts.

\begin{figure}[ht]
\centering
\begin{overpic}[width=0.45\textwidth]{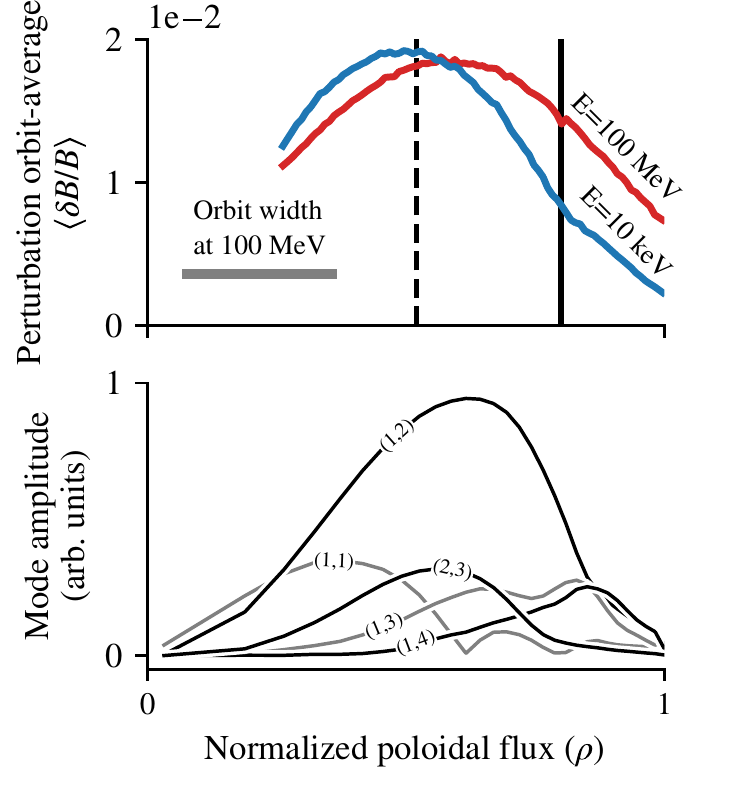}
\put(21,90){a)}
\put(21,48){b)}
\end{overpic}
\caption{
a) Mean perturbation amplitude along the particle trajectory for 10 keV and 100 MeV electrons in the fully stochastic JET case. 
Horizontal axis is the marker initial position at the outer mid-plane.
Dashed vertical line denotes the position from where the markers were launched when evaluating the transport coefficients.
The solid vertical line is the new loss surface that has been moved inwards from the edge.
The horizontal grey bar illustrates the orbit width of an 100 MeV electron.
b) Radial profiles of the dominant modes in the fully stochastic JET case.
The brackets $(n,\,m)$ are the toroidal and poloidal mode numbers of the mode.
}
\label{fig: profiles}
\end{figure}

The transport barrier appears because $\delta B /B$ is not uniform radially, see Fig.~\ref{fig: profiles}a, and the perturbation is significantly weaker at the edge.
However, this barrier is weaker for more energetic electrons because their orbit drift takes them closer to the core where the perturbation is stronger.
Therefore, the orbit-averaged value of $\delta B /B$ increases with energy for electrons at the edge.
This increases transport the effect being larger than the reduction due to FOW effects.
From Fig.~\ref{fig: profiles}b, we can observe that $\delta B /B$ in the fully stochastic case peaks where the dominant $(1,2)$ mode has a peak.
The modes $(1,3)$ and $(1,4)$ that have resonant surface at the edge are comparably lower which explains the drastic reduction in $\delta B /B$.

As for the ITER case, the orbit averaging is valid for the entire energy range.
At 100~MeV, the predicted decrease in transport is 10\% which is a significantly smaller reduction than what was observed in the simulations.
One should note that this case differs from all the others in that the perturbation is localised poloidally; $\delta B / B$ is strongest near the coils at the low-field side.
The theory assumes uniform perturbation, which might explain the difference seen here.

As the final point, we note that the numerically evaluated parallel correlation length in the artificial ITER cases is close to the estimate based on the toroidal mode number, and the perpendicular correlation lengths are comparable to the radial width of the modes.
In the JET cases, the dominant modes (recall Fig.~\ref{fig: profiles}b) have $n=1$, which gives an estimate $\lambda_\parallel\approx50$~m that is an order of magnitude higher than the numerically evaluated value.
On the other hand, the perpendicular correlation length is comparable to the mode width which is roughly half of the minor radius, i.e.~0.6~m.
Therefore, one can use the estimate $\lambda_\perp\sim$ mode width, in order to make an initial assessment whether FOW effects should be considered.

\section{Summary and conclusions}
\label{sec:conclusions}

While the orbit-following simulations broadly agree with the theoretical results in Ref.~\cite{Hauff_2009}, it was found that in disruptions magnetic field structures and non-uniform perturbation may dominate the  energy dependence of the transport.
In extreme cases, the transport of energetic particles ceases completely when their orbit crosses confined field line regions, or, somewhat unexpectedly, the transport increases above the transport level that would be observed if the particles would follow the field lines exactly.

The evaluation of the radial advection and diffusion coefficients was motivated by the prospect of including them in a reduced kinetic model, in order to capture the effect of the 3D magnetic field on the runaway electron dynamics.
Based on our results, if the magnetic field is spatially inhomogeneous, it is more appropriate to evaluate the transport coefficients numerically, instead of relying on the analytical model.

Simulation results differ from theory mostly in the orbit-averaging regime.
This was especially true in the case where the magnetic field was perturbed with external coils, which could be attributable to the perturbation being poloidally localised in this case.
Simulations done within the guiding center approximation yielded results that were practically identical to those obtained by tracing the complete gyromotion.
Only for $E>50$ MeV in some cases, the guiding center results showed higher transport.
We note that Ref.~\cite{Carbajal_2020} reported a difference between gyro-orbit and guiding center results, but this was due to particles being born inside the islands which was not considered here.

One of the motivations for investigating the energy dependence of the transport was the discrepancy where the RE transport observed in experiments was orders of magnitude lower than that predicted by the Rechester-Rosenbluth diffusion coefficient.
However, here  FOW effects were found to reduce the transport by an order of magnitude only at $E=100$~MeV, which is on the high end of plausible runaway electron energy spectra in ITER-like disruptions and is well beyond the energy of seed populations in any runaway scenario.
The energy reduction depends on the perpendicular correlation length of the magnetic perturbation, which was found to be of the same order as the radial width of the mode.
Importantly, when the stochastic field is caused by MHD activity, such that the mode widths are comparable to minor radius, and not due to microscale turbulence, the transport of REs is not reduced in the energy range of interest.

In our simulations the Rechester-Rosenbluth estimate fared well in the zero orbit-width limit except for one case, the one with major islands present, where it over-estimated the transport by two orders of magnitude.
We emphasize that this reduction was not because particles are born inside islands where they remain trapped, as was the case in Ref.~\cite{Carbajal_2020}, as in this case the markers where initialized within the stochastic region; instead the mere presence of the islands reduced the transport.

\ack
We acknowledge the CINECA award under the ISCRA initiative, for the availability of high performance computing resources and support.
The authors are grateful to  M.~Hoppe, P.~Svensson and I.~Pusztai for fruitful discussions.  This work was supported by the European Research Council (ERC-2014-CoG grant 647121), the Swedish Research Council (Dnr.~2018-03911), and the EUROfusion - Theory and Advanced Simulation Coordination (E-TASC). 
The work has been carried out within the framework of the EUROfusion Consortium and has received funding from the Euratom research and training programme 2014-2018 and 2019-2020 under grant agreement No 633053. 
The views and opinions expressed herein do not necessarily reflect those of the European Commission.

\appendix

\section{Transport coefficient evaluation}
\label{app:a}

We assume the radial transport of particles is given by the following advection-diffusion equation
\begin{equation}
\label{eq: fokker-planck}
\frac{\partial f}{\partial t} = -\frac{\partial }{\partial \rho} K f +\frac{\partial^2 }{\partial \rho^2} D f,
\end{equation}
where $f(\rho,t;\mu,E)$ is the distribution function, and $K(\rho;\mu,E)$ and $D(\rho;\mu,E)$ are the advection and diffusion coefficients, respectively.
The energy, $E$, and magnetic moment, $\mu$, are kept as parameters since they remain invariant in the transport process we consider.

The transport coefficients are likely to have a radial dependence because the magnitude of $\delta B/B$ may vary and any islands that are present affect the transport.
However, the radial dependence is neglected here and, instead, the transport coefficients are evaluated only at a single radial position.
At this position, we initialize a population of markers representing electrons.
The markers are initialized at the outer mid-plane and their toroidal coordinate is sampled from a uniform distribution.

The markers are traced with ASCOT5 and the transport coefficients are extracted from the results.
There are several ways this can be done~\cite{Sarkimaki_2016}, and here we choose to evaluate the coefficients based on the distribution of loss time, i.e. the time it takes for a marker from the beginning of the simulation to cross the separatrix for the first time.
Some markers might be radially confined, e.g.~inside islands, and, therefore, the marker population is simulated until this \emph{loss-time distribution} becomes saturated.

\begin{figure}[t]
\centering
\begin{overpic}[width=0.45\textwidth]{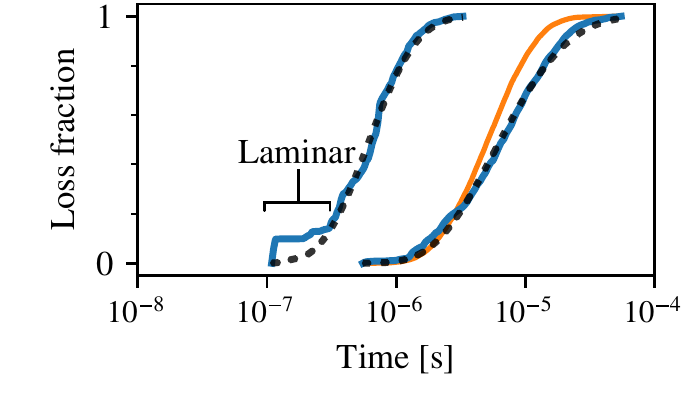}
\end{overpic}
\caption{
Illustration of the transport coefficient evaluation and loss times.
The cumulative first-passage-time distribution, Eq.~\eqref{eq: firstpassagetime}, is fitted (dotted black lines) to the cumulative particle loss-time distribution (solid blue lines), and the coefficients are obtained from the fit parameters.
Two cases are shown: the one on the left has also laminar losses which appear as ``steps" in the loss time distribution.
The orange line is the loss-time distribution where the coefficients obtained from the fit are used in a transport model, but employing a reflective inner boundary.
}
\label{fig: loss time verification}
\end{figure}

The transport coefficients are extracted from the loss-time distribution by fitting an analytical model to the results; however, the analytical model is based on several simplifications.
When one assumes that the transport is uniform radially (i.e.~$K$ and $D$ do not depend on position), the advection term is positive (i.e.~the flow is towards the edge), and the inner boundary extends to minus infinity
\footnote{
In reality, the inner boundary is reflective and it is located at the magnetic axis or at the boundary between the confined and stochastic field-line regimes.}
, then the \emph{cumulative} loss-time distribution obeys the so-called \emph{first-passage time distribution}~\cite{Domin_1996} which, in probability theory is known as the inverse Gaussian distribution.
The corresponding cumulative distribution is
\begin{align}
\label{eq: firstpassagetime}
T(t) = &\Phi \left({\frac {1 }{\sqrt {2Dt}}}\left(Kt-\Delta\rho\right)\right) \nonumber\\
&+\exp \left({\frac {K\Delta\rho }{D }}\right)\Phi \left(-{\frac {1 }{\sqrt {2Dt}}}\left(Kt+\Delta \rho\right)\right),
\end{align}
where $\Phi$ is the cumulative normal distribution.
The first-passage time refers to the time instant when a random walker launched from a given position first passes through a pre-defined coordinate (separated by a distance $\Delta\rho$), which exactly is the case here when the simulations measure the time when a marker crosses the separatrix for the first time.
In addition to fitting, the moments of the loss-time distribution can be used to estimate the transport coefficients as $K=\Delta \rho / \operatorname{Mean}[t]$ and $D=(\Delta \rho)^2 \operatorname{Var}\left[t\right] / 2 \operatorname{Mean}[t]^3$, although the results are more susceptible to noise.

Even though the approximations behind Eq.~\eqref{eq: firstpassagetime} are drastic, the formula fits the data well: at any time the difference in losses deviate less than 10~\% from the simulation results.
Two example fits are shown in Fig.~\ref{fig: loss time verification}: the one on the left (faster losses) is for 100 MeV electrons ($\xi=0.9$) and the one on the right is for 10 keV in the fully stochastic JET case.

The data on the left has a step-like structure at the beginning indicating that multiple markers are lost at the same time.
In this \emph{laminar transport} particles are lost before they have decorrelated, and the advection-diffusion model does not describe this early-time behaviour.
Still, the trend is captured by the fit even though the details are not.

For the second case the shape of the fit is slightly different than the data, which is due to the actual transport not being uniform in $\rho$.
We can assess what effect omitting the reflective inner boundary in Eq.~\eqref{eq: firstpassagetime} has, by including one in Eq.~\eqref{eq: fokker-planck} and solving the equation numerically.
We use the coefficients obtained from the fit (second case), an initial radial distribution that is identical to the one used in the orbit-following simulations, and set the reflecting boundary right next to the marker initial position.
The cumulative loss-time distribution obtained this way (the orange line) deviates from the earlier results as particles are lost faster due to the reflection.
The take-away message here is that since a reflective boundary is inherently present in the data to which Eq.~\eqref{eq: firstpassagetime} is fitted, we can expect that the coefficients we obtain overestimate the actual transport.

Finally, the advection is negative only in the presence of islands but islands are not present in most cases studied in this work.

\section{Correlation length evaluation}
\label{app:b}

The (auto)correlation length is defined as
\begin{equation}
\label{eq: correlation length}
\lambda \equiv \int_0^\infty C(l) dl,
\end{equation}
where $C(l)$ is the (auto)correlation function of the magnetic field perturbation:
\begin{equation}
\label{eq: correlation fun}
C(l) \equiv 
\frac{\left<
\int_{-\infty}^\infty\tilde{B}(r)\tilde{B}(r+l)dr
\right>}{\left<
\int_{-\infty}^\infty\tilde{B}^2(r)dr
\right>},
\end{equation}
where $\tilde{B}$ is the perturbation component that is perpendicular to the unperturbed field. 
The brackets are averages over all realizations of $\tilde{B}$.

In most cases the perturbation is not isotropic; in this paper, for example, the perturbation potential oscillates along the field lines.
Therefore, it is customary to consider the parallel and perpendicular correlation lengths and functions separately.
For the parallel correlation length (function), the integrals in Eqs.~\eqref{eq: correlation length}~--~\eqref{eq: correlation fun}, are performed along the field lines.

The correlation function peaks at $l=0$ due to the convolution in the numerator (the denominator is just a normalization factor).
It is generally assumed that, in both directions, $C(l)$ has the form of Eulerian correlation function:
\begin{equation}
\label{eq: eulerian correlation}
C(l) \sim 
\exp\left(-\frac{l^2}{2\lambda_\parallel^2}\right),
\end{equation}
but this is valid only if $\tilde{B}$ is a Gaussian random noise.

\begin{figure}[ht]
\centering
\begin{overpic}[width=0.45\textwidth]{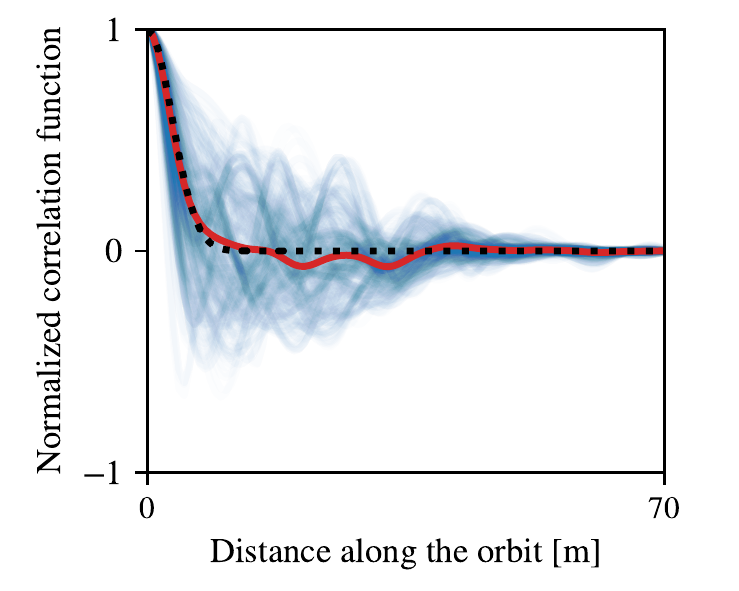}
\end{overpic}
\caption{
Numerically evaluated magnetic field (parallel) correlation function.
Horizontal axis is the distance in meters.
The blue lines are the correlation functions calculated for individual field lines, and the red line is the ensemble average of these, i.e.~the actual correlation function in Eq.~\eqref{eq: correlation fun}.
The black line is the Eulerian correlation function, Eq.~\eqref{eq: eulerian correlation}, where $\lambda_\parallel$ was obtained by integrating the actual correlation function (the red line) according to Eq.~\eqref{eq: correlation length}.
}
\label{fig: correlation fun}
\end{figure}

We evaluate the parallel correlation function numerically by choosing a radial position, and then finding the field-line $(R,\phi,z)$ coordinates in the axisymmetric field.
The integrals in Eq.~\eqref{eq: correlation fun} are evaluated in the perturbed field along these coordinates, after which we move the toroidal coordinate by a random angle, and repeat the evaluation.
The process is repeated until the ensemble average has converged and we obtain the correlation function.
Figure~\ref{fig: correlation fun} illustrates how the numerically evaluated correlation function compares to the estimate in Eq.~\eqref{eq: eulerian correlation}.

%Diffusive processes in a stochastic magnetic field
%Plasma transport in stochastic magnetic fields. Part 3. Kinetics of test particle diffusion

\section*{References}
\bibliographystyle{iopart-num}
\bibliography{energyscan}

\end{document}